\documentclass[aps,twocolumn,pra]{revtex4}

\usepackage[normalem]{ulem}%    % strickthrough the text \sout{text to be stricked} 
\usepackage{newlfont}
\usepackage{color}         % color to the text \textcolor{red,green,blue}{text to be colored}
\usepackage{amssymb}
\usepackage{amsfonts}
\usepackage{amsmath}
\usepackage{wasysym}
\usepackage{graphicx}
\usepackage{epsfig}
\usepackage{amsthm}
\usepackage{bm}
\usepackage{mathrsfs}
\usepackage{epstopdf}
\usepackage{theorem}
\usepackage{multirow}
\usepackage{slashbox}

\usepackage[colorlinks=true, citecolor=blue, urlcolor=blue ]{hyperref}
\begin{document}

\title{Universality in Distribution of Monogamy Scores for Random Multiqubit Pure States}

\author{Soorya Rethinasamy\(^{1,2}\), Saptarshi Roy\(^1\), Titas Chanda\(^{1,3}\), Aditi Sen(De)\(^1\), Ujjwal Sen\(^1\)}

\affiliation{\(^1\)Harish-Chandra Research Institute, HBNI, Chhatnag Road, Jhunsi, Allahabad 211 019, India}
\affiliation{\(^2\)Birla Institute of Technology and Science, Pilani, Rajasthan 333031, India}
\affiliation{\(^3\) Instytut Fizyki im. Mariana Smoluchowskiego, Uniwersytet Jagiello\'nski,  \L{}ojasiewicza 11, 30-348 Krak\'ow, Poland.}
%, and\\
%Homi Bhaba National Institute, Anushaktinagar, Mumbai - 400094, India}

\begin{abstract}
Monogamy of quantum correlations provides a way to study restrictions on their sharability in multiparty systems.  We find the critical exponent of these measures,  above which randomly generated multiparty pure states satisfy the usual monogamy relation, and show that the critical power decreases with the increase in the number of parties.  For three-qubit pure states, we  detect that W-class states are \emph{more} prone to being nonmonogamous as compared to the GHZ-class states. We also observe a different criticality in monogamy power up to which random pure states remain nonmonogamous. We prove that  the ``average monogamy" score asymptotically approaches its maximal value on increasing the number of parties. Analyzing the monogamy scores of random three-, four-, five- and six-qubit pure states, we also report that almost all random pure six-qubit states  possess maximal monogamy score, which we confirm by evaluating statistical quantities like mean, variance and skewness of the distributions. In particular, with the variation of number of qubits, means of the distributions of monogamy scores for random pure states approach to unity -- which is the algebraic maximum -- thereby conforming to the known results of random states having maximal multipartite entanglement in terms of geometric measures.

%Monogamy of quantum correlations (QC) paves the way to study restrictions on the distribution of QC in multiparty systems.  Some fundamental properties of monogamy are investigated using monogamy score for measures from both entanglement based and information theoretic paradigm. The critical monogamy power of these measures, i.e., the power of QC measure in the monogamy score,  above which all randomly generated multiparty pure states exhibit monogamy are computed. For three-qubit pure states, we quantitatively detect that W-class states are `more' nonmonogamous compared to the GHZ-class states by evaluating the `average nonmonogamy fraction' of these classes of states. The distribution of monogamy scores are investigated for random three, four, five and six-qubit pure states. 
%These distributions are analyzed statistically and certain universal features are reported. We have shown analytically, for almost all measures of QC, the average monogamy score approach its maximal value on increasing the number of parties. We conform to the above result by looking at the transition of the distribution of monogamy score towards the maximal value on continually increasing the system size from three to six. Finally, we propose a connection between multiparty entanglement and monogamy score.
\end{abstract}

\maketitle

\section{Introducion}

%general introduction to entanglement and general QC
Quantum entanglement \cite{qent}, one of the most striking features in quantum mechanics,  is the essential resource \cite{ent_resource} for a plethora of quantum information protocols like quantum teleportation \cite{teleportation}, quantum dense coding \cite{dense_coding}, entanglement-based quantum cryptography \cite{crypto_monogamy}, one-way quantum computation \cite{oneway} etc. These protocols revolutionize the existing communication and computation schemes based on laws of classical mechanics.
 Due to immense  importance of entanglement, over the years, several detection methods like partial transposition  \cite{ppt1} based on positive maps \cite{ppt2}, entanglement witness, \cite{wit1, wit2, wit3}, and quantifiers such as distillable entanglement \cite{dis}, entanglement of formation \cite{eof, concurrence}, logarithmic negativity \cite{neg,logneg} have been proposed.
 On the other hand, it has also been realized that quantum mechanical systems can exhibit nonclassical phenomena which cannot be explained by using the theory of entanglement, and hence a different resource theory has been developed where unlike separable states, ``classically correlated" states in the computational basis are the free states \cite{discord2}. These measures, independent of entanglement, belonging to a fine-grained paradigm of quantum correlations (QC), have their origin in the concepts of information theory or geometry of states or thermodynamics.  Examples of such measures include quantum discord \cite{discord1}, geometric quantum discord \cite{geoqd}, quantum work deficit \cite{wd1,wd2}, and quantum deficit \cite{deficit} (see review \cite{discord2}).
% Apart from entanglement, quantum mechanical systems can also posses quantum correlations which are beyond entanglement and are definitely non-classical. These can be captured by measures like quantum discord \cite{discord1,discord2} and quantum work deficit \cite{wd1,wd2}, which owe their inception from information theoretic concepts. Therefore, in general, the measures to quantify quantum correlations (QC) include both from the entanglement separability paradigm like concurrence \cite{concurrence}, negativity \cite{neg}, logarithmic-negativity \cite{neg,logneg}, Entanglement of formation \cite{eof} etc. and from the information theoretic genre like quantum discord and work deficit.

%Intro to monogamy
Beyond the bipartite domain, understanding of QC, even for pure states shared by multiple parties, 
is limited due to its complex structure. 
%is a long standing problem in quantum information science. 
%Towards the aim of quantifying multipartite QC, an important direction which has been evolved both for pure and mixed states is based on the distance metric \cite{geoent} and the corresponding measure, commonly known as the geometric measure of QC, is defined as the distance between a given state and the set of free states of the respective resource theory. Recent studies \cite{entropy_bounds, winter, eisert, qcomp2} showed that random pure states having moderate number of parties possess a high amount of geometric measure of multipartite entanglement and hence they are useful for quantum information processing tasks like quantum teleportation \cite{teleportation}, although they are shown to be disadvantageous for the classical processing part of  one-way quantum computation.
%
In this paper,  
%we choose another notable avenue to seek for features  of QC for random multipartitie pure states. In particular, 
we study the distribution of QC among the various parties of a random multipartite quantum state, with the help of the concept of ``monogamy" \cite{review}. Unlike classical correlations, which can be shared freely between different parties, monogamy of QC \cite{review, ent_monogamy,CKW} restrains arbitrary sharing of QC among the various parties of a multipartite quantum state. Specifically, in a tripartite scenario, if two parties share maximal QC, they cannot at all be quantum  correlated with the third party. The quantitative version of this concept, in the form of an inequality, was formulated by Coffman, Kundu, and Wootters \cite{CKW} involving bipartite QC measures for arbitrary states having varying number of parties. 
%Application of the concept of monogamy
Apart from its fundamental importance \cite{ent_monogamy, foundation-mono,foundation-mono2,foundation-mono3,foundation-mono4,foundation-mono6, mono-other4}, such notion turned out to have applications in diverse areas of physics including quantum crypography \cite{mono_cryp,mono_app3}, identification of quantum channels \cite{mono_app4}, and many body physics \cite{mono_app1,mono_app2,mono_app5}. 
The monogamy inequality can also give rise to a quantity, known as monogamy score \cite{monogamyscore}, which can take both negative as well as non-negative values and can be used for characterization of QC. We ask here the following question: Do non-negative monogamy scores for random multiparty pure states behave in a fashion that is similar to the known results obtained for  geometric measure of entanglement (cf. \cite{eisert, winter})? 
%\textcolor{red}{The monogamy score can be both non-negative and negative.} 
%In this paper, we address the question whether \textcolor{red}{in situations where the monogamy score is non-negative,} it can be argued as a valid multipartite QC measure for states with high number of parties, irrespective of the choice of \textcolor{red}{bipartite QC measure to construct the monogamy score}, thereby mimicking the results obtained for geometric measures of entanglement. 
We answer it affirmatively, for a large spectrum of bipartite QC measures, required to construct the monogamy score.
%This provides motivation for foundational studies to get a deeper understanding of the concept of monogamy of QC and indeed a lot of columns \cite{foundation-mono,foundation-mono2,foundation-mono3,foundation-mono4} have been spent fruitfully in this direction. 

%previous works in this direction
%\noindent \textcolor{red}{Write a section about previous works in this direction with lots of references!}

%What we did in this paper
%In this paper, we choose another notable avenue in which we consider to study the monogamy properties of randomly chosen multiparty pure states via the monogamy score \cite{monogamyscore}. 
Most of the computable measures are known to be nonmonogamous for three-qubit pure states while their squares often satisfy the monogamy relations \cite{CKW, mono-other1,mono-other2,mono-other3,mono-other4, negativitymonogamysq, dis_sq}. For a given QC, the power (exponent) that makes  all three qubit pure states to be monogamous often lies between $1$ and $2$. An exception is quantum work deficit \cite{foundation-mono}. We first find out the critical power required for obtaining monogamy for all three-qubit pure states. We observe that  criticalities are different for the two classes inequivalent with respect to stochastic local operations and classical communication (SLOCC)-viz. the GHZ- and the W-classes \cite{Dur_SLOCC}. 
%Such study reveals a surprising fact that for a finite range of values in powers between $1$ and $2$ of QC measures, all randomly generated pure three-qubit states violate monogamy relation with all known computable QC quantifier.

When the above analysis is repeated for randomly generated pure states of four-, five- and six-qubits, we find that the critical value in power of the bipartite QC measures in the monogamy score decreases with the increase in the number of parties. Moreover, we observe that for entanglement measures, the percentage of states to exhibit nonmonogamy jumps to a lower value when the powers of entanglement measures are strictly greater than $0$ (upto numerical accuracy). 

%We then consider the original question by considering monogamy score with the QC power higher than the criticalities, so that the monogamy score is always non-negative for random pure states.
 We finally show that the monogamy score approaches its algebraically allowed maximum value on increasing the number of qubits for random pure states irrespective of the QC measure. We then examine several statistical quantities like mean, variance and skewness of the distributions of monogamy scores for random pure states of three-, four-, five- and six-qubits. Specifically, such investigations  reveal that there is a transition in the distribution of monogamy score from lower mean values to higher mean values, and from being right-skewed to left-skewed  with the increase in the number of qubits from three to six. Therefore, we conclude that the monogamy scores based on any quantum correlation measure  reaches their maximum value for most of the random pure states with moderate number of parties, thereby supporting an analytical result (in the asymptotic limit) obtained in this paper and mimicking the pattern of multiparty geometric measure of entanglement. 

%Any state with a non-negative monogamy score is termed as monogamous. Firstly, we make a comparative study of the critical monogamy power of different QC measures above which all states of the two SLOCC inequivalent classes (GHZ and W) \cite{Dur_SLOCC} of 3-qubit pure states becomes monogamous. The above analysis is repeated for randomly generated pure states of four, five and six qubits. We also find the critical value of the monogamy power below which all states of the above classes exhibit nonmonogamous nature.
%For related studies involving monogamy power and monogamy relations see \cite{mono-other1,mono-other2,mono-other3,mono-other4}.  
% Then we move on to the actual distribution of monogamy scores for various measures and perform statistical analysis to extract out various universal (measure independent) characteristics  of the considered distributions. We analytically show, the monogamy score approaches its maximum value on incresing the number of qubits irrespective of the QC measure. The transition of the distribution of monogamy score to its maximal value is demonstrated by increasing the number of qubits from three to six. Finally, we show, distribution of monogamy score mimics the properties of geometric measure of entanglement and try to establish a relationship between monogamy score and multiparty entanglement. 

%Organisation of the paper
The paper is organized as follows. We discuss the concept of monogamy score in Sec. \ref{sec:ms}. Subsequently, the criticalities in the power of bipartite measures with respect to their monogamy scores are investigated in Sec. \ref{sec:cri}. A comparative study between the criticalities of the GHZ- and the W-class states are given in Sec. \ref{sec:cri_ghz_w},  followed by a similar analysis for random pure states with higher number of qubits in Sec. \ref{sec:cri_mul}. We then shift our focus to the actual distributions of monogamy score in Sec. \ref{sec:statistics} and extract universal features of these distributions by computing their statistical properties. We then study the difference in the distributions for the GHZ- and W-class states in Sec. \ref{sec:stat_ghz_w} and then move on to a higher number of parties in Sec. \ref{sec:stat_mul}. 
%We propose a connection between monogamy score and multiparty entanglement in \ref{sec:stat_ent}. 
We present concluding remarks in Sec. \ref{sec:conclusion}.

%\begin{enumerate}
%\item Universal feature of monogamy score to mimic the properties of multiparty entanglement measures.
%
%\item 
%In this manuscript, we show some universal characteristics of monogamy scores defined with respect to various quantum correlation measures when the power of monogamy is above the critical value for certain specific classes of random pure states
%\end{enumerate}

\section{Monogamy scores}
\label{sec:ms}
The monogamy score \cite{ent_monogamy,CKW, monogamyscore}, for an arbitrary \(N\)-party quantum state, $\rho_{12\ldots N}$, with respect to a bipartite quantum correlation measure, $\cal Q$,
%The multiparty quantum correlation measure, 
 with the first party \cite{othernodal} as the ``nodal" observer, is defined as
\begin{equation}
\delta_{\cal Q}\equiv \delta_{\cal Q}(\rho_{12\ldots N}) = {\cal Q}_{1:{\text rest}} - \sum_{i = 2}^N {\cal Q}_{1:i},
\label{eq:def_ms}
\end{equation}
where \({\cal Q}_{1:{\text rest}} \equiv {\cal Q}(\rho_{1:{\text rest}})\) is the quantum correlation in the 1:rest bipartition and  \( {\cal Q}_{1:i} \equiv {\cal Q}(\rho_{1:i})\) denotes the quantum correlation $\mathcal{Q}$ between the nodal party and the $i^{\text{th}}$ party  of the reduced density matrix, $\rho_{1i}$ $(i=2,3,...,N)$ of $\rho_{12...N}$. 
%In this case, we consider the first party as the nodal observer. 
%By considering other parties as nodal one, one can  similarly obtain a valid monogamy score.  
%One can consider another party as the nodal observer as well.
\(\delta_{\cal Q} (\rho_{12\ldots N}) \geq 0\) implies that the  state, \(\rho_{12\ldots N}\), is monogamous with respect to \({\cal Q}\), and otherwise it is nonmonogamous. 
%Therefore, for a multiparty state, satisfying monogamy relation for $\mathcal{Q}$, means ${\cal Q}_{1:{\text rest}} \geq \sum_{i = 2}^N {\cal Q}_{1:i}$.

In this paper, the chosen measures of quantum correlation (QC) from the entanglement-separability paradigm are concurrence,  entanglement of formation, negativity, and logarithmic negativity, while
%All of these satisfy the criterion of ``good" measures of quantum correlation as prescribed in \cite{limit_ent}.  
 from the information theoretic genre, we opt for quantum discord and quantum work deficit. The measures are chosen due to their analytical or numerical computability. For definitions of the above measures, see Appendix \ref{appendix}.

\section{Criticalities in monogamy power}
\label{sec:cri}
It is known that there exist three- or more-party quantum states which violate the monogamy relation for certain QC measures. At the same time, it was proven that if one considers squares of several of these measures, 
 such states become monogamous \cite{CKW, negativitymonogamysq, dis_sq}. QC measures, exhibiting this behavior include all the computable measures considered in this paper except quantum work-deficit \cite{foundation-mono}.
 % for which higher power of these measures are required to satisfy monogamy for three qubit pure states as also considered in this paper in succeeding sections.
 
 We address here the question whether instead of an integer power, QC measures raised to  positive real numbers are enough to satisfy the monogamy relation for three-qubit pure states and for states with a higher number of parties. Specifically, we consider 
\begin{eqnarray}
\delta_{\mathcal{Q}^\alpha} = \mathcal{Q}^\alpha_{1:\text{rest}} - \sum_{i=2}^{N} \mathcal{Q}^\alpha_{1:i},
\end{eqnarray} 
  for some positive real number, $\alpha$, which typically lies between $1$ and $2$. In this respect, one should note that if $\mathcal{Q}$ is a valid measure of QC,  $\mathcal{Q}^\alpha$, where $\alpha$ is a positive real number, is also a valid measure. 
Moreover, notice that if 
%for two-qubit states, $0 \leq\mathcal{Q}\leq 1$, and hence apart from tailor-made systems, typically for all $i$, 
$\mathcal{Q}(\rho_{1:i}) < 1$ and if $\mathcal{Q}(\rho_{1:\text{rest}}) < 1$,  we have $\delta_{\cal Q^{\alpha \rightarrow \infty}}\rightarrow 0$ for arbitrary multiqubit states. It implies that in the asymptotic limit ($\alpha \rightarrow \infty$), all states become monogamous.  Therefore it is interesting to examine whether there exists a finite value of $\alpha = \alpha_c$, such that for all  pure states  $\delta_{\mathcal{Q}^\alpha} \geq 0$, for $\alpha \geq \alpha_{c}$. Let us call $\alpha_c$ as the critical monogamy power. Furthermore, for $\alpha \rightarrow 0$, it trivially follows that $\delta_{\mathcal{Q}^{\alpha\rightarrow0}} \rightarrow -N + 2$, for an $N$-party state, which reduces to $-1$, for three-qubit states, implying violation of monogamy inequality for all states, independent of $\mathcal{Q}$. Let us now define a quantity which quantifies the number of Haar uniformly generated states, violating monogamy relation for a given $\mathcal{Q}$, among all multiparty random pure states. We call it as the fraction of nonmonogamous states, $f$. Mathematically, for a given $\mathcal{Q}$ and $\alpha$,
\begin{eqnarray}
f = \frac{\text{number of nonmonogamous states}}{\text{total number of randomly generated states}}.
\end{eqnarray}
in the limit of a large total number of states generated.
When $\alpha \rightarrow 0$, $f \rightarrow 1$ for all multiparty random pure states.

%Such studies of finding the critical value of the monogamy power, $\alpha^c$, have been studied for various classes of states [ref] and also for generalized monogamy relations in [ref].

 % i.e., all states are nonmonogamous for any $\mathcal{Q}$ raised to the power $\alpha$, when $\alpha \rightarrow 0$. 
Assuming continuity, one can expect that in the neighborhood of $\alpha = 0$, $f$ remains close to unity. Interestingly, however, our analysis reveals that the fraction of nonmonogamous states, $f$, remain frozen to unity even for some finite values of $\alpha$ leading to another criticality of $\alpha$. We denote this critical value of $\alpha$ upto which the value of $f$ stays unity as $\alpha_p$.
  
%  for a small region near $\alpha =0$, the fraction of nonmonogamous states for some given class might be pinned to unity. However, during our analysis we find that, some classes of states are completely nonmonogamous even upto some finite values of $\alpha$. So, another criticality to look at is the value of $\alpha$ upto which the fraction of nonmonogamous state remains frozen to unity. We denote this critical value of $\alpha$ to be $\alpha_p$.

%Then if we are considering monogamy relation with respect to $\mathcal{Q}$, then it is also worthwhile to contemplate monogamy relations with respect to the $\alpha^{\text{th}}$ power of $\mathcal{Q}$.

\begin{figure}[h]
\includegraphics[width=\linewidth]{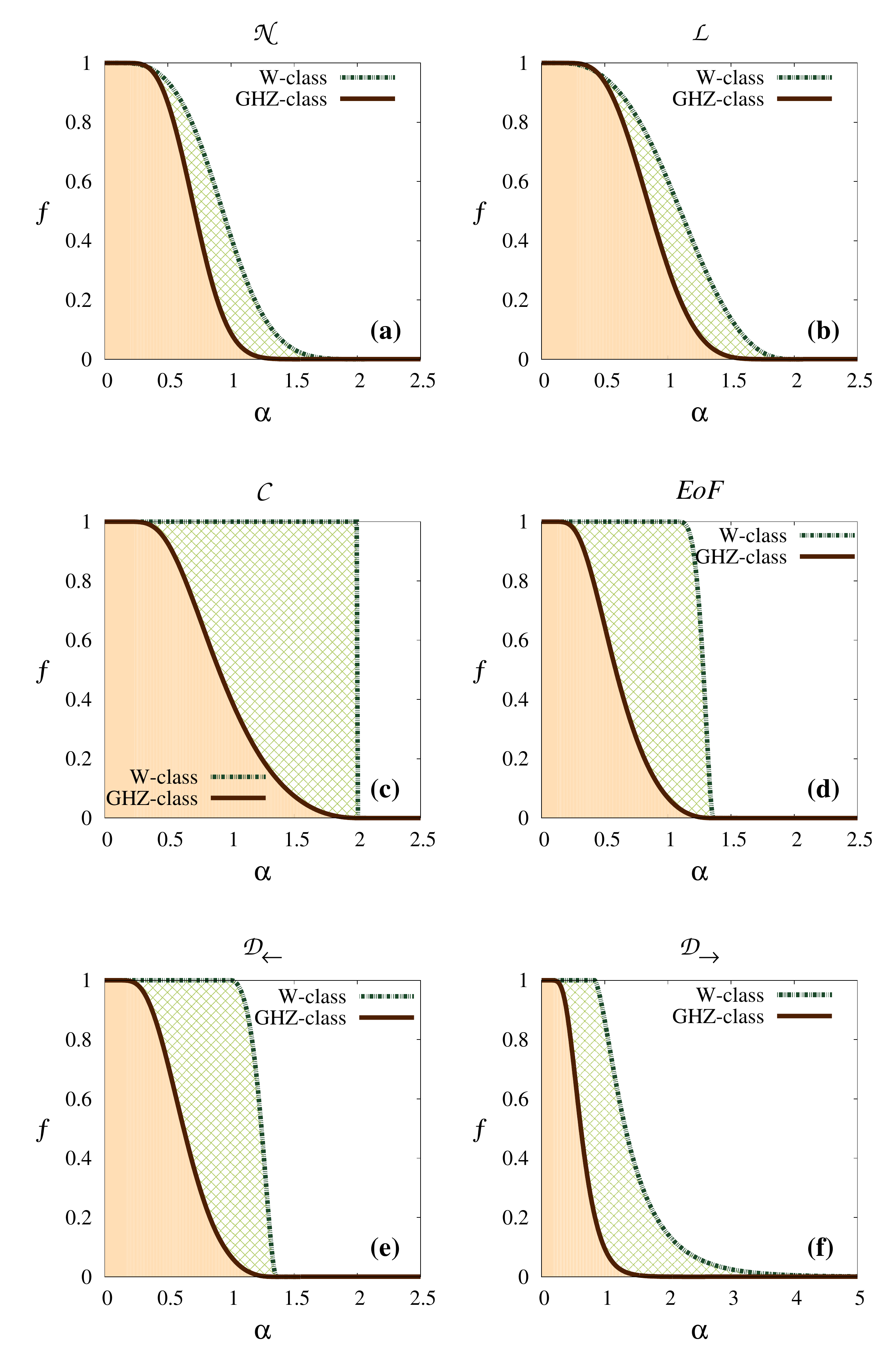}
\caption{(Color online.) Trends in fraction of nonmonogamous states, $f$, against monogamy power, $\alpha$, of QC measures for the GHZ- and the W-class states. The figures labelled (a)-(f) show  $f$, for negativity ($\mathcal{N}$), logarithmic negativity ($\mathcal{L}$), concurrence ($\mathcal{C}$), entanglement of formation ($EoF$), quantum discord by measurement in the second party ($\mathcal{D}_\leftarrow$) and in the first party ($\mathcal{D}_\rightarrow$) of a quantum state. The shaded regions under each curve gives the area under the nonmonogamy fraction denoted as $\mathcal{M}$ (see Eq. \eqref{eq:avg_nonmonogamy_fraction}). All axes are dimensionless.}
\label{fig:cri_ent}
\end{figure}

 In this section, we investigate both the criticalities, $\alpha_c$ as well as $\alpha_p$, for both entanglement-based  and information-theoretic measures of QC. Specifically: 
\begin{enumerate}
\item We Haar uniformly generate $10^6$ three-qubit states \cite{Dur_SLOCC} and perform a comparative study between the $\alpha_p$ and $\alpha_c$ values for different classes of three-qubit pure states. 

\item To investigate the effect of the  increase in the number of parties on $\alpha_c$, we also randomly generate multiqubit pure states having four, five, and six parties. 
%\item We evaluate $\alpha_c$ values for multi-qubit random pure states ($4,5$ and $6$ qubits) and study the effct of increasing the number of qubits on the critical monogamy power, $\alpha_c$. 

%\item We define and compute the \emph{average monogamy fraction}, which  quantifies the  nonmonogamous nature of a given class of states.
\end{enumerate}
%We investigate these aspects in the succeeding sections.

\subsection{Comparative study between GHZ- and W-class states}
\label{sec:cri_ghz_w}
\begin{figure}[h]
\includegraphics[width=\linewidth]{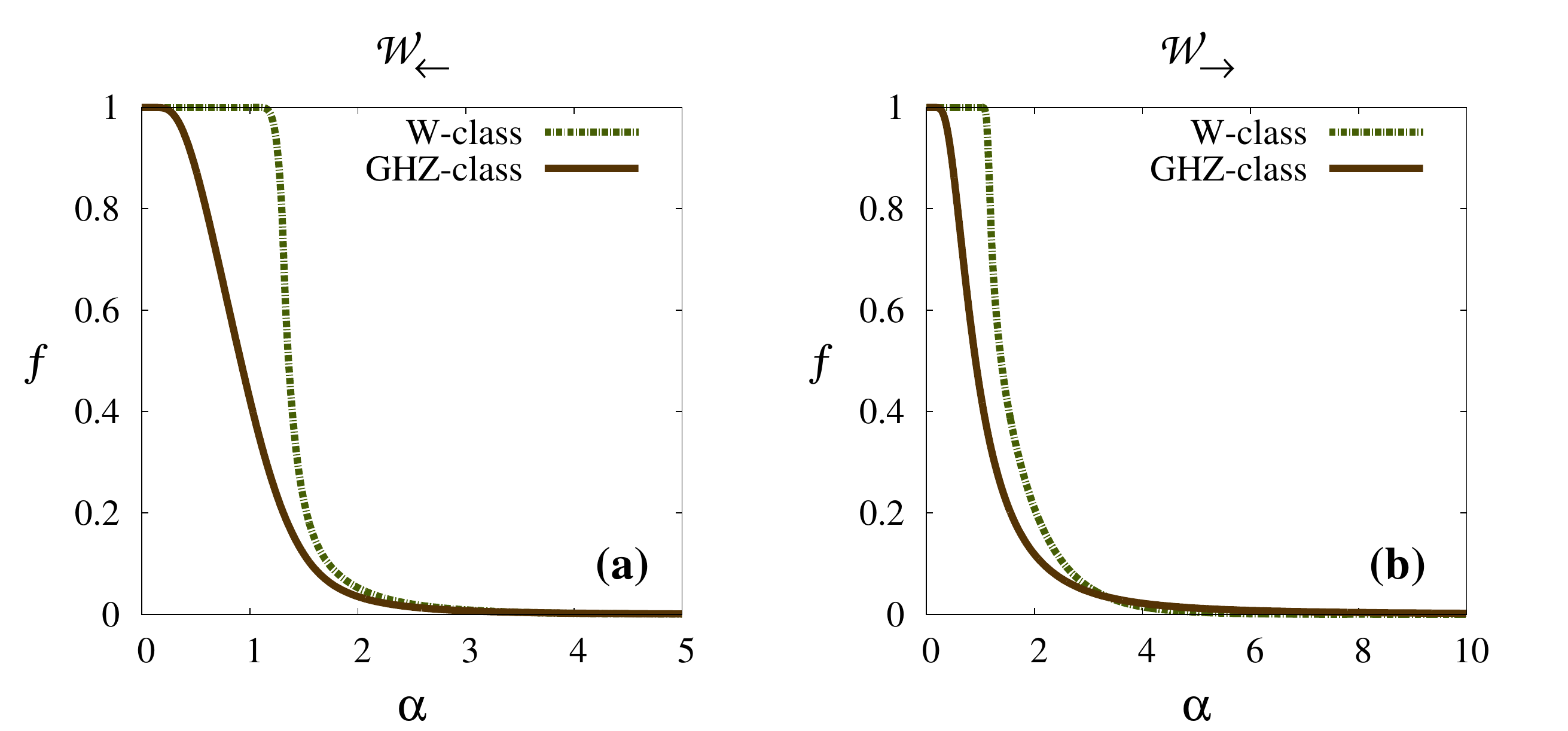}
\caption{(Color online.) Behavior of fraction of nonmonogamous states ($f$) vs. monogamy power ($\alpha$) for quantum work deficit when measurement is in (a) the second ($\mathcal{W}_\leftarrow$) and (b) the first part ($\mathcal{W}_\rightarrow$) of the $A:BC$, $A:B$, and $A:C$ bipartition of a given three-qubit pure state, $|\psi\rangle_{ABC}$.   
% Criticalities in the monogamy power for quantum work deficit. The figures labelled (a) and (b) show the fraction of nonmonogamous states, f, for work deficit right ($\mathcal{W}_\leftarrow$) and work deficit left $\mathcal{W}_{\rightarrow}$ respectively.  Note that $\mathcal{W}_{\leftarrow}$ and $\mathcal{W}_{\rightarrow}$ denote the quantum work deficit after measurement in the second and the first party respectively. 
All axes are dimensionless.}
\label{fig:wd_cri}
\end{figure}

The set of three-qubit pure states consists of two disjoint subsets belonging to two SLOCC inequivalent classes, namely the GHZ- and the W-class \cite{Dur_SLOCC}. Note that the states from the W-class is a set of measure zero and hence in the generation of random three-qubit pure states, W-class states cannot be  found, thereby requiring an independent simulation for the W-class states.
%We randomly generate pure tripartite states Haar uniformly as described in Ref. \cite{Karolbook} to investigate Figs.  \ref{fig:Histo_neg_m} and \ref{fig:Histo_neg_m_2copy}. 
%We quote the results of the critical value of the monogamy power for these classes of states for both the entanglement-based and information theoretic measures. Figs. \ref{fig:cri_ent} and \ref{fig:wd_cri} show the fraction of nonmonogamous states, f, for increasing values of $\alpha$ for various measures of QC. We find qualitatively different features for the GHZ and W-class states. Specifically, from numerical analysis and Figs.  \ref{fig:cri_ent} and \ref{fig:wd_cri} we find,
We Haar uniformly generate three-qubit pure states randomly from both the classes and compare their respective $\alpha_c$ and $\alpha_p$ for both entanglement and information-theoretic measures. Our analysis reveals that the states from the GHZ- and the W-class show qualitatively distinct behavior with respect to criticalities which are independent of the choice of the QC measures.
We find (see Figs.  \ref{fig:cri_ent} and \ref{fig:wd_cri}, and Table. \ref{table:critical} :

\begin{enumerate}
\item For a large set of values of $\alpha$, the fraction of nonmonogamous states for the GHZ-class is less than that of the W-class, i.e., $f^{\text{GHZ}}_{\mathcal{Q}^\alpha} \leq f^{\text{W}}_{\mathcal{Q}^\alpha}$, for any quantum correlation measure $\mathcal{Q}$.
% Here, f is the fraction of nonmonogamous states obtained for a given value of $\alpha$. This is the reminiscent of the fact 
It implies that W-class states are more nonmonogamous than the GHZ-class states.

\item The value of $\alpha_p$ obtained for the GHZ-class, denoted by $\alpha_p^{\text{GHZ}}$ is lower than that of the W-class, $\alpha_p^{\text{W}}$, i.e., $\alpha_p^{\text{GHZ}} \leq \alpha_p^{\text{W}}$. 

\item Like $\alpha_p$, the critical value of $\alpha$ for which all states become monogamous is always smaller for the GHZ-class states, $\alpha_c^{\text{GHZ}} \leq \alpha_c^{\text{W}}$, where $\alpha_C^{\text{GHZ}}$ and $\alpha_c^{\text{W}}$ denote critical values of $\alpha$ for which $\delta_{\mathcal{Q}^\alpha} \geq 0$ for the GHZ- and the W-class respectively. This result can be interpreted as the consequence of $f^{\text{GHZ}}_{\mathcal{Q}^\alpha} \leq f^{\text{W}}_{\mathcal{Q}^\alpha}$.

%\item \textcolor{red}{In the case of concurrence and entanglement of formation, all randomly chosen W-class states show nonmonogamous nature up till some particular finite value of $\alpha = \alpha_p$. Therefore, $f^{\text{W}} = 1$ for all values of $\alpha \leq \alpha_p$ for the above mentioned measures. 
%Note that $f$ falls below $1$ for $\alpha > \alpha_p$ and hits $0$ when $\alpha = \alpha_c$.}

\item For concurrence, we know that $\delta_{\mathcal{C}^2} = 0$ for all W-class states. We find that $\alpha_p = \alpha_c = 2$ for all states from the W-class. It implies that all values of $\alpha < \alpha_p = \alpha_c = 2$, we have $f = 1$, i.e., all randomly chosen W-class states are nonmonogamous with respect to $\mathcal{C}^\alpha$ (i.e., $\delta_{\mathcal{C}^\alpha}<0$) for $\alpha < 2$.  Moreover, $f$ has a finite jump from $1$ to $0$ at $\alpha = 2$, which indicates the known result of vanishing three-tangle for the W-class states \cite{Dur_SLOCC}.

\item Interestingly, for the W-class states,  $\alpha_c$ of $\mathcal{L}$ and $\mathcal{C}$ coincide, at $\alpha_c = 2$, implying that no finite real values of $\alpha$ exist between $0$ and $2$ to make all states monogamous. Such characteristic is also seen for the states from the GHZ-class in case of monogamy relation of concurrence.

\item We also observe that $\alpha_p$ is reasonably high for the W-class states in case of concurrence ($\mathcal{C}$), entanglement of formation ($EoF$), quantum discord ($\mathcal{D}_\leftarrow$ and $\mathcal{D}_\rightarrow$), and work-deficit ($\mathcal{W}_\leftarrow$ and $\mathcal{W}_\rightarrow$). In all these cases, $\alpha \gtrsim 1$.% or $\alpha \approx 1$.

% we find that $\alpha^c_{\text{W}} \geq \alpha^c_{\text{GHZ}}$.

\end{enumerate}

The above picture remains qualitatively similar for both entanglement-based and information theoretic QC measures. However, there are some differences also. In particular:
\begin{enumerate}
\item For quantum work deficit, we find that, $\alpha_c > 10$ independent of the party on which measurements are performed. To our knowledge, this is the only known measure of QC which requires such a high value of $\alpha$ to satisfy monogamy relation. 
%\textcolor{red}{Exact $\alpha_c$ values wud b put after rechecking.}

\item In case of information-theoretic measures, both $\alpha_c$ and $\alpha_p$ depend on the party where the measurement is performed. For example, in case of quantum discord, the value of $\alpha_c$ is drastically high when the measurement is carried out on the first party (see Table. \ref{table:critical}). 
%The critical values for the measures considered in the manuscript are listed in Table. \ref{table:critical}.
\end{enumerate}
%for work deficit, $\alpha_c > 10$ for measurement in either of the parties. Moreover, as expected, the results of the information theoretic measure depends on the party on which the measurement is performed. For example, in the case of quantum discord, the value of $\alpha_c$ is drastically different when the measurement is done on the first party (see Table. \ref{table:critical}). The critical values for the measures considered in the manuscript are listed in Table. \ref{table:critical}.

 A detailed analysis of the critical monogamy power of random pure states will be important for arguing monogamy score as a valid multiparty QC measure from the perspective of the properties of random pure states.
 % and a necessary condition for claiming this resemblance is the non-negativity of the the monogamy score which is guaranteed only when $\alpha \geq \alpha_c$.

%\textcolor{red}{A table of $\alpha_{p/c}$}

\begin{table}[h]
\begin{center}
\begin{tabular}{|c|c|c|c|}
    \hline
    Measure &Class &$\alpha_p$ &$\alpha_c$ \\
    \hline
    \multirow{2}{*}{Negativity ($\mathcal{N}$)}&GHZ &0.1467 &1.6735\\ \cline{2-4} 
        &W &0.0991 &1.9885\\
    \hline
    \multirow{2}{*}{Logarithmic negativity ($\mathcal{L}$)}&GHZ &0.1497 &1.8540\\ \cline{2-4} 
        &W &0.0991 &2\\
    \hline
    \multirow{2}{*}{Concurrence ($\mathcal{C}$)}&GHZ &0.1470 &2\\ \cline{2-4} 
        &W &2 &2\\
    \hline
    \multirow{2}{*}{Entanglement of formation ($EoF$)}&GHZ &0.0866 &1.3520\\ \cline{2-4} 
        &W &1.0410 &1.4280\\
    \hline
    \multirow{2}{*}{Quantum discord (left) ($\mathcal{D}_{\rightarrow}$)}&GHZ &0.1163 &3.4317\\ \cline{2-4} 
        &W &0.8382 &9.3492\\
    \hline
    \multirow{2}{*}{Quantum discord (right) ($\mathcal{D}_{\leftarrow}$)}&GHZ &0.0968 &1.3520\\ \cline{2-4} 
        &W &0.9797 &1.3608\\
    \hline
    \multirow{2}{*}{Work deficit (left) ($\mathcal{W}_{\rightarrow}$)}&GHZ &0.1183 &$>$10\\ \cline{2-4} 
        &W &0.9630 &$>$10\\
    \hline
    \multirow{2}{*}{Work deficit (right) ($\mathcal{W}_{\leftarrow}$)}&GHZ &0.0989 &$>$10\\ \cline{2-4} 
        &W &0.9799 &$>$10\\
    \hline
\end{tabular}
\caption{Comparative study of values of the two criticalities, $\alpha_p$ and $\alpha_c$ (see main text) for the GHZ- and the W-class states. }
\label{table:critical}
\end{center} 
\end{table}

%\begin{figure}[h]
%\includegraphics[width=\linewidth]{info_critical2}
%\caption{Criticalities in the monogamy power for information theoretic measures. Measurement is performed in the first party.}
%\end{figure}

Let us now introduce a measure to quantify the nonmonogamous nature of a given class of  multipartite states with respect to a given QC measure, $\mathcal{Q}$ for all $\alpha \leq \alpha_c$. It is defined  as the area under the curve of the fraction of nonmonogamous states with respect to the monogamy power of the considered quantum correlation measure $\mathcal{Q}$, which  
%We call this quantity as the \emph{average nonmonogamy fraction} ($\mathcal{M}$). 
for a given class of states, reads as
\begin{eqnarray}
\mathcal{M}_{\mathcal{Q}} = \int_{0}^{\alpha_c} f_{\mathcal{Q}^\alpha} \text{ d}\alpha ,
\label{eq:avg_nonmonogamy_fraction}
\end{eqnarray}
 provided $\alpha_c$ is finite.
%The average nonmonogamy score for the measures of quantum correlation considered in this manuscript are given in 
See  Table. \ref{table:avg_non_ms} for $\mathcal{M}_\mathcal{Q}$ values for different QC measures. Note that $\mathcal{M}_\mathcal{Q}$ is a single number
%This gives a quantative estimation of the well known fact that  states from the W-class are more nonmonogamous compared to the GHZ-class states for various measures of quantum correlation.
which composes information of both $\alpha_c$ and $\alpha_p$. As already argued before, $\mathcal{M}_\mathcal{Q}$ also shows that irrespective of the choice of $\mathcal{Q}$, on average, random pure states from the W-class are more nonmonogamous than that from the GHZ-class. 

\textbf{Remark:} Since the $\alpha_c$-value is very large for work deficit ($>10$),  $\mathcal{M}_{\mathcal{W}_{\leftarrow}}$
and  $\mathcal{M}_{\mathcal{W}_{\rightarrow}}$
%average nonmonogamy fraction
 are not numerically discernible in this case. 

\textbf{Note:} In the remainder of the paper, whenever we refer to quantum discord or work deficit, the measurement will be performed in the second party.
\begin{table}[h]
\begin{center}
\begin{tabular}{|c|c|c|}
    \hline
    Measure &Class & $\mathcal{M}_{\mathcal{Q}}$ \\
    \hline
    \multirow{2}{*}{Negativity ($\mathcal{N}$)}&GHZ &0.7245 \\ \cline{2-3} 
        &W &0.9441 \\
    \hline
    \multirow{2}{*}{Logarithmic negativity ($\mathcal{L}$)}&GHZ &0.8765 \\ \cline{2-3} 
        &W &1.0957 \\
    \hline
    \multirow{2}{*}{Concurrence ($\mathcal{C}$)}&GHZ &0.9498 \\ \cline{2-3} 
        &W &2.0000 \\
    \hline
    \multirow{2}{*}{Entanglement of formation ($EoF$)}&GHZ &0.6182 \\ \cline{2-3} 
        &W &1.2751 \\
    \hline
    \multirow{2}{*}{Quantum discord (left) ($\mathcal{D}_{\rightarrow}$)}&GHZ &0.6669 \\ \cline{2-3} 
        &W &1.4870 \\
    \hline
    \multirow{2}{*}{Quantum discord (right) ($\mathcal{D}_{\leftarrow}$)}&GHZ &0.6472 \\ \cline{2-3} 
        &W &1.2333 \\
    \hline
%    \multirow{2}{*}{Work deficit left}&GHZ &$\approx$1.1933 \\ \cline{2-3} 
%        &W &$\approx$1.6604 \\
%    \hline
%    \multirow{2}{*}{Work deficit right}&GHZ &X \\ \cline{2-3} 
%        &W &X \\
%    \hline
   \end{tabular}
\caption{$\mathcal{M}_\mathcal{Q}$ for various quantum correlation measures, $\mathcal{Q}$, for three-qubit pure states.}
 \label{table:avg_non_ms}
\end{center}
\label{tab:multicol}
\end{table}

\begin{figure}[h]
\includegraphics[width=\linewidth]{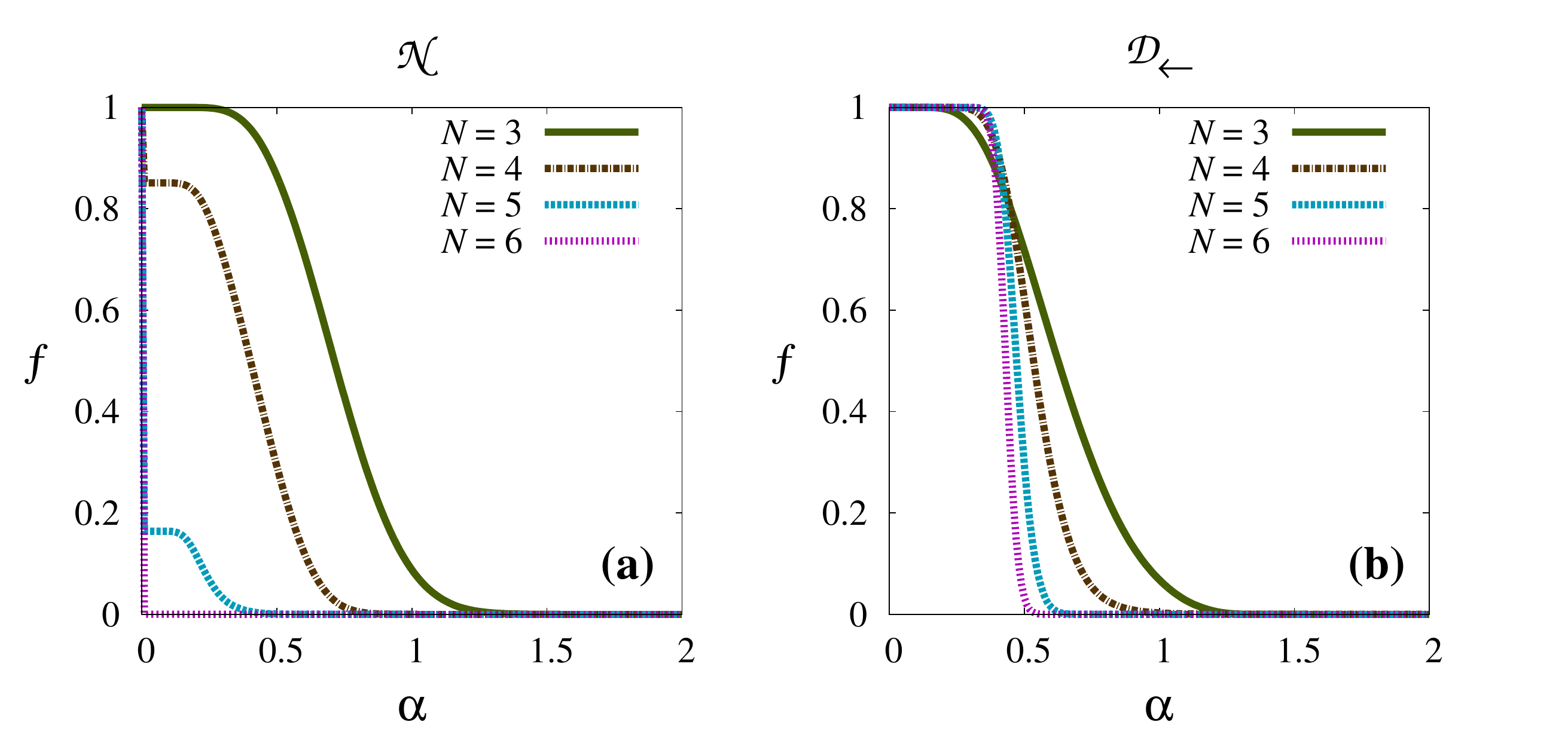}
\caption{(Color online.) Plot of $f$ against $\alpha$ with different number of parties, $N$. In (a), negativity is chosen as an entanglement measure, while in (b), $\mathcal{D}_\leftarrow$ is plotted. 
%Note that other entanglement measures also display identical features. In (b) $\mathcal{D}_\leftarrow$ is chosen for similar study. 
Comparing (a) and (b), we note that unlike entanglement measures, $f$ for information-theoretic measures do not show any jump with $\alpha \rightarrow 0+$. 
%Criticalities in the monogamy power for increasing number of qubits. In (a) we show for negativity ($\mathcal{N}$), how the fraction of nonmonogamous states changes on increasing the number of parties. Although we show this for negativity, other entanglement based QC measures display identical features.  The same thing for discord right is depicted in (b).  
All axes are dimensionless.}
\label{fig:mul_cri}
\end{figure}

\subsection{Uniformity in criticalities for multiqubit states}
\label{sec:cri_mul}

%\textcolor{green}{Computing criticalities in $\alpha$ for states with higher number of parties; special reference to Dicke states.}
Let us now move on to the computation of criticalities in $\alpha$ for random pure states of higher number of qubits. Specifically, we  focus on the change in the value of $\alpha_c$ for a given $\mathcal{Q}$ with increasing number of parties, $N$. We observe the emergence of a universal feature of $\alpha_c$ with respect to $\mathcal{Q}$ when the number of qubits is varied. However, there is a stark contrast in the response of $f$ and $\alpha_p$ obtained  for entanglement and information-theoretic measures with the increase in the number of parties. Let us now enumerate the similarities and differences of $\delta_{\mathcal{Q}^\alpha}$ by varying the number of qubit, $N$, from $3$ to $6$.

\begin{enumerate}
\item The value of $\alpha_c$ and $\mathcal{M}_\mathcal{Q}$ decreases for all $\mathcal{Q}$ with the increase of $N$. See Fig. \ref{fig:mul_cri} for some typical QC measures. Similar traits can also be seen for other QC measures. 

%\item The average nonmonogamy fraction nonmonogamous states progressive
\item Although $\alpha_c$ shows some similarities, the fraction of nonmonogamous states demonstrates contrasting behavior depending on the choices of the measure. In particular, in case of entanglement measures, $f$ starts from $1$ at $\alpha = 0$ and then jumps to a value which is strictly smaller than $1$ with $\alpha \neq 0$, thereby causing a discontinuity in $f$ with $\alpha$. Such discontinuity in $f$ is not observed in case of information theoretic QC measures.

%\item For all entanglement based measures, on increasing the number of parties (qubits) from three to four, the fraction of nonmonogamous states make a sudden jump from unity to a lower value on increasing the monogamy power $\alpha$ from zero.

\item With the increase of $N$, the value of $\alpha_p$ for entanglement measures decreases, while it is increasing for information theoretic ones. We observe this feature for moderate number of parties.

\end{enumerate}

%The above observations hint towards the fact that as we go on increasing the number of parties, the average nonmonogamy fraction monotonically decreases for all $\mathcal{Q}$, i.e., 
The above observations indicate that for moderate  number of qubits, all the random pure states become monogamous with respect to arbitrary QC measures. We will analyze this in greater detail in the succeeding sections and examine the transition from nonmonogamy to monogamy with the variation of the number of parties. 
%Furthermore, a natural question which also comes up is how the monogamy scores are actually distributed for various measures and varying number of parties. 

\section{Statistics of monogamy scores for Multipartite random Pure States}
\label{sec:statistics}
%\textcolor{red}{A short intro required.}
 %In this section, we focus on the distribution of monogamy scores for various classes of states with respect to different measures of quantum correlation. 
 In recent years, it has been argued, by using the concept of geometry of quantum states \cite{Karolbook}, that random pure multipartite states with a moderate number of parties are highly entangled \cite{entropy_bounds, winter, eisert, qcomp2}, and hence  they can be resourceful in quantum information processing tasks, especially in quantum communication protocols. In this section, we investigate the distribution of monogamy scores of random pure multiqubit states for different QC measures. 
 The patterns of distribution of monogamy scores show some universal characteristics irrespective of the measures used to probe them. We first report the difference between the distribution patterns of the GHZ- and the W-class states. We then move on to  a higher number of parties and comment how the corresponding spreads change when we consider  random pure states with more than three parties. The results obtained in this regard, is in good agreement with the previous results based on geometric entanglement measures.  
 %We finally propose a connection between monogamy scores with different kinds of QC measures and genuine multiparty entanglement measures.

\subsection{Distribution of monogamy scores for multipartite QC measures of three-party states: GHZ vs. W class states}
\label{sec:stat_ghz_w}
To carry out the investigation of three-qubit pure states, we generate $10^6$ pure states $|\psi\rangle_{ABC}$,  Haar uniformly from both the GHZ- and the W-classes (see Appendix \ref{appendix}). 
%The distribution of monogamy scores are examined for several quantum correlations from both entanglement based and information theoretic genres. 
%These measures are briefly discussed in Appendix \ref{appendix}.
 The analysis is performed by considering the entanglement and information-theoretic  measures as well as their squares. We are interested to scrutinize the qualitative characteristics of monogamy scores for $\mathcal{Q}^\alpha$, i.e. $\delta_\mathcal{Q}^\alpha$  in $\alpha<\alpha_c$ and $\alpha>\alpha_c$ regions by considering certain statistical quantities, viz. mean, variance, and skewness \cite{stat_book}  defined as
\begin{eqnarray}
\text{Mean: }\mu &=& \langle \delta_{\mathcal{Q}^\alpha} \rangle, \nonumber \\
\text{Variance: }\sigma^2 &=& \langle \delta_{\mathcal{Q}^\alpha}^2 \rangle - \mu^2 \nonumber, \\ \text{Skewness: }\kappa &=& \Big\langle \Big(\frac{\delta_{\mathcal{Q}^\alpha} -\mu}{\sigma} \Big)^3 \Big\rangle.
\label{eq:statistical_def}
\end{eqnarray}
%We focus on the properties of genuinely entangled tripartite states. This includes both GHZ and W class states. \textcolor{red}{What do u mean by taking together? Do we gain anything extra?.} $10^6$ states each of GHZ and W were generated and the statistics of the monogamy scores of the various measures are reported below. For quantum discord, and work deficit, the measurements are performed on the 2nd subsystem.
\begin{figure}[h]
\includegraphics[width=\linewidth]{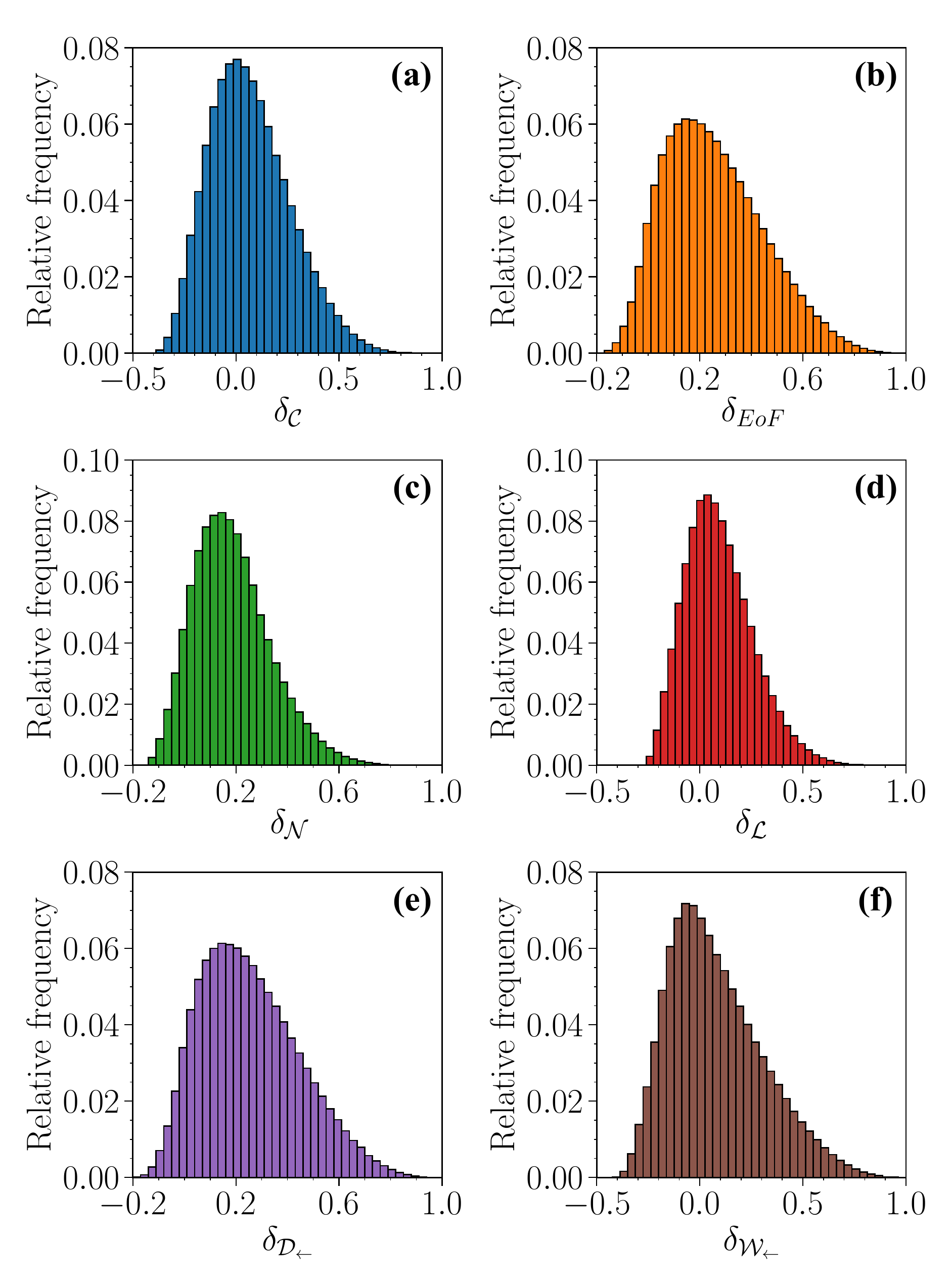}
\caption[Monogamy Score]
{(Color online.) Relative frequency of monogamy scores for random three-qubit states. Measures chosen are  (a) concurrence (b) entanglement of formation (c) negativity (d) logarithmic negativity (e) quantum discord (right) and (f)  quantum work deficit (right). Note that  Haar uniform simulation of three-qubit random pure states always belong to the GHZ-class. The shape of the distribution remains qualitatively similar if one considers $\delta_{\mathcal{Q}^2}$ as also seen from Table. \ref{table:GHZ and W Statistical Parameters squared}.  All axes are dimensionless.}\label{fig:Monogamy Scores for GHZ States}
\end{figure}
\begin{table}[h]
\begin{center}
\begin{tabular}{|c|c|c|c|c|}
    \hline
%    $\mathcal{Q}$ & Class & \textbf{Mean} & $\sigma^2 = \langle \delta_{\mathcal{Q}}^2 \rangle - \mu^2$ & $\gamma = \langle \big(\frac{\delta_{\mathcal{Q}}-\mu}{\sigma}\big)^3 \rangle$ \\
$\mathcal{Q}$ & Class & \textbf{Mean} & \textbf{Variance} & \textbf{Skewness} \\
    \hline
    \multirow{2}{*}{$\mathcal{N}$} & GHZ & 0.18542 & 0.022174 & 0.62577 \\ \cline{2-5} 
        & W & 0.025438 & 0.0069282 & 0.38597 \\
    \hline
    \multirow{2}{*}{$\mathcal{L}$}&GHZ & 0.094092 & 0.026725 & 0.59757\\ \cline{2-5} 
        &W &-0.023887 & 0.01132 & 0.15495 \\
    \hline
    \multirow{2}{*}{$\mathcal{C}$}&GHZ &0.068952 & 0.037962 & 0.48944\\ \cline{2-5} 
        &W &-0.19631 & 0.0089384 & -0.076814\\
    \hline
    \multirow{2}{*}{$EoF$}&GHZ &0.25496 & 0.036393 & 0.50209 \\ \cline{2-5} 
        &W &-0.062687 & 0.0022365 & -0.66105\\
    \hline
    \multirow{2}{*}{$\mathcal{D}_{\leftarrow}$}&GHZ & 0.25496 & 0.036393 & 0.50209\\ \cline{2-5} 
        &W & -0.062687 & 0.0022365 & -0.66105\\
    
    \hline
    \multirow{2}{*}{$\mathcal{W}_{\leftarrow}$}&GHZ &0.079392 & 0.051408 & 0.64816\\ \cline{2-5} 
        &W &-0.085636 & 0.0033781 & -0.9464\\
    \hline
\end{tabular}
\caption{Mean, variance and skewness values for the GHZ- and the W-class states for various  QC measures.}
\label{table:GHZ and W Statistical Parameters}
\end{center} 
\end{table}
\begin{table}[h]
\begin{center}
\begin{tabular}{|c|c|c|c|c|}
    \hline
%    $\mathcal{Q}^2$ & Class & $\mu = \langle \delta_{\mathcal{Q}^2} \rangle$ & $\sigma^2 = \langle \delta_{\mathcal{Q}^2}^2 \rangle - \mu^2$ & $\gamma = \langle$ $\big($\tiny{$\frac{\delta_{\mathcal{Q}^2} -\mu}{\sigma}$} \normalsize $\big)^3\rangle$ \\
$\mathcal{Q}^2$ & Class & \textbf{Mean} & \textbf{Variance} & \textbf{Skewness} \\
    \hline
    \multirow{2}{*}{$\mathcal{N}^2$} & GHZ & 0.413269 & 0.027109 & 0.19832 \\ \cline{2-5} 
        & W & 0.14462 & 0.015031 & 0.90635 \\
    \hline
    \multirow{2}{*}{$\mathcal{L}^2$}&GHZ & 0.37581 & 0.023849 & 0.40478 \\ \cline{2-5} 
        &W &0.13071 & 0.010451 & 0.76232 \\
    \hline
    \multirow{2}{*}{$\mathcal{C}^2$}&GHZ &0.33335 & 0.034293 & 0.50371\\ \cline{2-5} 
        &W & 0 & - & -\\
    \hline
    \multirow{2}{*}{$EoF^2$}&GHZ &0.3985 & 0.040395 & 0.32512 \\ \cline{2-5} 
        &W &0.060853 & 0.0036504 & 1.1312\\
    \hline
    \multirow{2}{*}{$\mathcal{D}^2_{\leftarrow}$}&GHZ & 0.42308 & 0.040715 & 0.23827\\ \cline{2-5} 
        &W & 0.074828 & 0.0043115 & 0.84245\\
    
    \hline
    \multirow{2}{*}{$\mathcal{W}^2_{\leftarrow}$}&GHZ &0.31472 & 0.043941 & 0.46134\\ \cline{2-5} 
        &W &0.058729 & 0.003126 & 0.70358\\
    \hline
\end{tabular}
\caption{Mean, variance and skewness values for the GHZ- and the W-class states for the  squares of various QC measures.}
\label{table:GHZ and W Statistical Parameters squared}
\end{center} 
\end{table}
%\begin{table}[h]
%\centering
%\begin{tabular}{|c|c|c|c|}
%\hline
%$\mathcal{Q}$ & $\mu = \langle \delta_{\mathcal{Q}} \rangle$ & $\sigma = (\langle \delta_{\mathcal{Q}}^2 \rangle - \mu^2)^{1/2}$ & $\gamma = \langle \delta_{\frac{\mathcal{Q}-\mu}{\sigma}}^3 \rangle$  \\\hline
%$\mathcal{C}$ & 0.068952 & 0.037962 & 0.48944 \\\hline
%$\mathcal{N}$ & 0.18542 & 0.022174 & 0.62577  \\\hline
%$\mathcal{E}$ & 0.25496 & 0.036393 & 0.50209 \\\hline
%$\mathcal{L}$ & 0.094092 & 0.026725 & 0.59757 \\\hline
%$\mathcal{D}_\leftarrow$ & 0.25496 & 0.036393 & 0.50209 \\\hline
%$\mathcal{W}_\leftarrow$ & 0.079392 & 0.051408 & 0.64816 \\\hline
%\end{tabular}
%\caption{Statistical Parameters of $\delta_\mathcal{Q}$ for GHZ states}
%\label{table:1}
%\end{table}
\begin{figure*}[ht]
\includegraphics[width=0.75\linewidth]{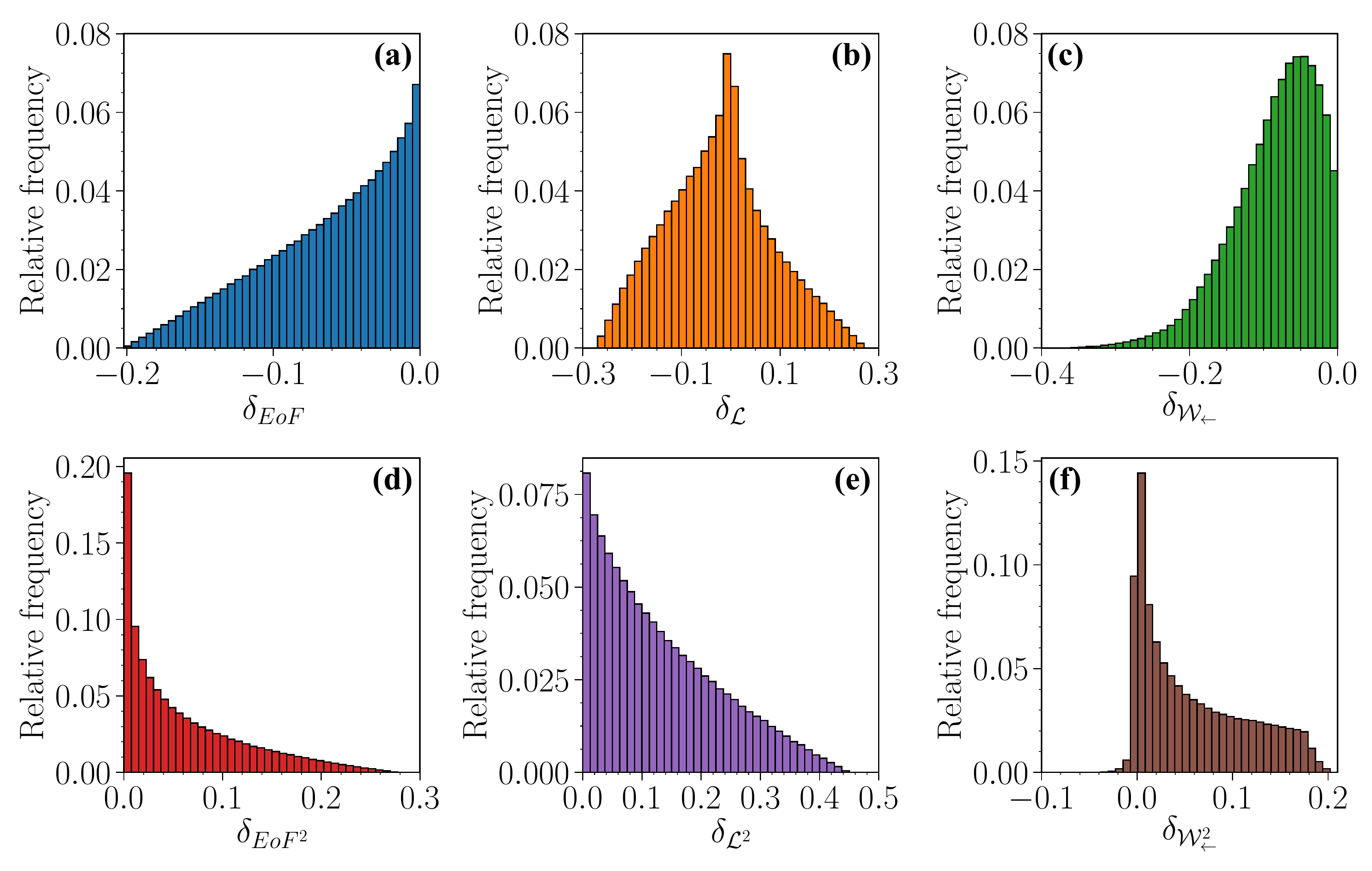}
\caption[Monogamy Score]
{(Color online.) Relative frequency of $\delta_\mathcal{Q}$ of states from the W-class  (a) $EoF$  (b) $\mathcal{L}$ and (c) $\mathcal{W}_\leftarrow$. In (d)-(f), similar measures are chosen for studying the distribution of $\delta_{\mathcal{Q}^2}$. Unlike the GHZ-class, the distribution of $\delta_{\mathcal{Q}}$ are quite contrasting to $\delta_{\mathcal{Q}^2}$ (see Tables. \ref{table:GHZ and W Statistical Parameters} and \ref{table:GHZ and W Statistical Parameters squared}). All axes are dimensionless.}\label{fig:Monogamy Scores for W States}
\end{figure*}
%In this section we look at the differences between the GHZ and the W class. 
%\textcolor{red}{We propose a quantum game, where a shopkeeper presents a black box which always produces GHZ states or W States, but we don't know which. Given a large enough sample size, could we distnguish between the two based purely on the statistical measures of the various scores?}

The key traits that emerge from the inspection of the GHZ- and the W-class states are summarized below:
\begin{enumerate}
\item The first important feature, we notice, is the universality in the distribution of monogamy scores for various QC measures in case of random pure states, belonging to the GHZ-class. In particular, we find that the mean of the distribution of $\delta_\mathcal{Q}$ \textcolor{red}{are} always positive for all $\mathcal{Q}$, as well as it is positively skewed leading to the Bell typed distribution tilted towards zero. Quantitatively, the $\sigma^2_{\delta_\mathcal{Q}}$ and $\kappa_{\delta_\mathcal{Q}}$ values tend to lie within  small windows, viz. $0.02 < \sigma^2_{\delta_\mathcal{Q}} < 0.05$ and $0.48 < \kappa_{\delta_\mathcal{Q}} < 0.65$. See Table. \ref{table:GHZ and W Statistical Parameters}. 
  %for this class of states and even quantitatively the skewness and variance of $\delta_\mathcal{Q}$ share some similar trends like $0.02 < \sigma^2_{\elta_\mathcal{Q}} < 0.05$ and $0.48 < \gamma_{\delta_\mathcal{Q}} < 0.64$. \textcolor{red}{Is the last info abt the amt of skewness or variance is needed? We can just say positively skewed and quantitatively, the $\sigma^2_{\delta_\mathcal{Q}}$ and $\gamma_{\delta_\mathcal{Q}}$ values tend to lie within specific ranges. See table.}
%The mean of the scores of all the measures is positive for the GHZ class, which is not the case in the latter.

%\item Furthermore GHZ states have positive skewness for the monogamy score distributon of all measures whereas the W class shows no such trend. This type of trend is also present in the measure value in the A:BC bipartition. GHZ class states have predominantly higher values and left skewness for all measures, but W class has a lower mean and both positive and negative skewness.
\item On the other hand, random pure states chosen from the W-class do not show any qualitatively similar pattern. For example, we observe that mean and skewness of $\delta_\mathcal{Q}^{\text{W}}$ do not even have the same sign for different QC measures (see Table. \ref{table:GHZ and W Statistical Parameters} and Fig. \ref{fig:Monogamy Scores for W States}). 

\item Furthermore, our analysis reveal that for the GHZ-class states, distributions of $\delta_{\mathcal{Q}^2}^{\text{GHZ}}$ and $\delta_{\mathcal{Q}}^{\text{GHZ}}$  are qualitatively similar (see Tables. \ref{table:GHZ and W Statistical Parameters} and \ref{table:GHZ and W Statistical Parameters squared}). However, for W-class states, a contrasting behavior in the distributions of $\delta_{\mathcal{Q}}^{\text{W}}$ and $\delta_{\mathcal{Q}^2}^{\text{W}}$ are observed. See Fig. \ref{fig:Monogamy Scores for W States}.
%Some comments on using squared QC measures. How results from power 1 case get altered?}   
\end{enumerate}  
As it is already known from \cite{Dur_SLOCC}, an easy test to distinguish two inequivalent classes in three-qubit pure states is to find the tangle ($\delta_{\mathcal{C}^2}$). Vanishing tangle guarantees the W-class states while the non vanishing value of $\delta_{\mathcal{C}^2}$ ensures the GHZ-class states.
 %score for all the states, and if are all zero, we can comment with confidence that the box produces W-class states. However, our 
 The above analysis provides a course-grained picture of the properties that are typical to all the QC measures, and are not just restricted to any specific one.

\begin{figure*}[t]
\includegraphics[width=\textwidth]{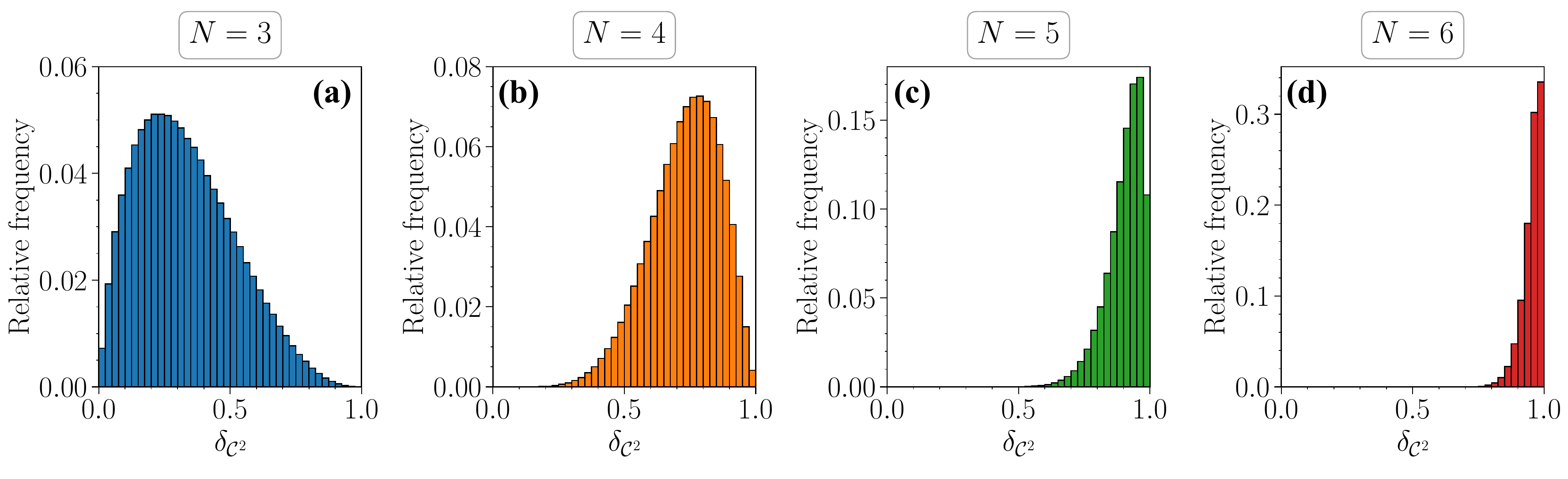}
\caption[ConcSqParties]
{(Color online.) Patterns of distribution of monogamy scores for squared concurrence for randomly chosen pure states with (a) three-, (b) four-,   (c)five- and (d) six-qubits. Figure (d) confirms that most of the random six-qubit pure states possess high values of $\delta_{\mathcal{C}^2}$. All axes are dimensionless.}\label{fig:ConcSqParties}
\end{figure*}

\begin{figure*}[t]
\includegraphics[width=\linewidth]{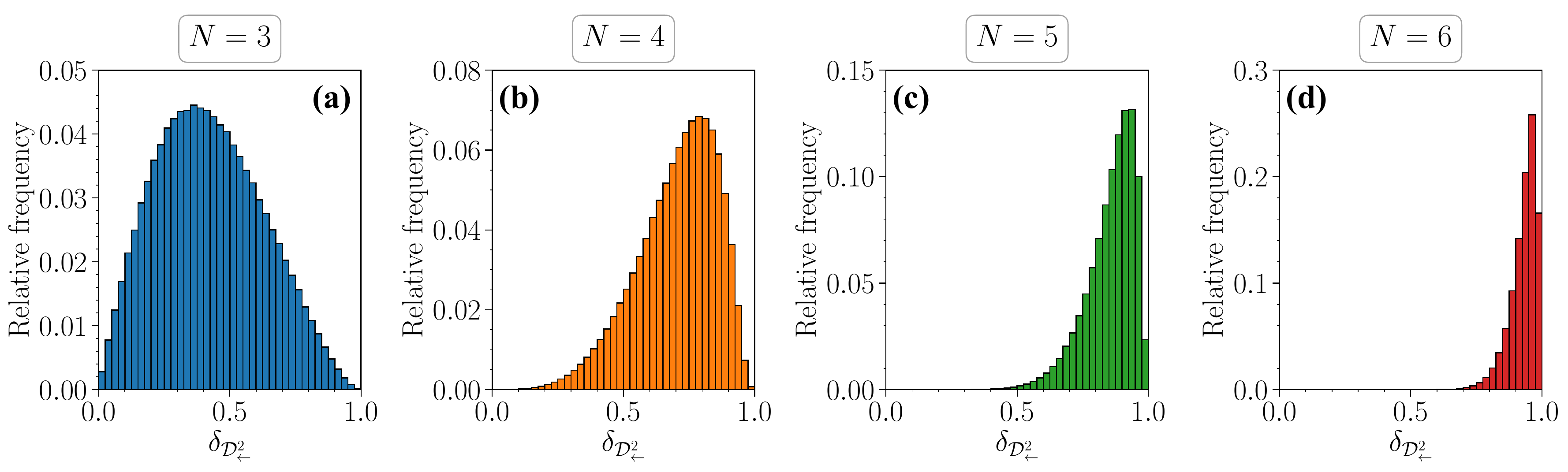}
\caption[DiscSqParties]
{(Color online.) Trends of distribution of $\delta_{\mathcal{D}_\leftarrow^2}$ for different values of $N$ ((a)-(d) 3,4,5 and 6 qubits). Note that $\delta_{\mathcal{C}^2}$ and $\delta_{\mathcal{D}_\leftarrow^2}$ behave in a similar fashion and their behavior is same as multiparty geometric entanglement measures as reported in \cite{eisert, qcomp2}.  All axes are dimensionless.}\label{fig:DiscSqParties}
\end{figure*}

\subsection{Monogamy scores mimic properties of multipartite entanglement measures}
\label{sec:stat_mul}
In the preceeding section, we observe that all the distribution of monogamy scores for random three-qubit pure states which typically belong to the GHZ-class, follow a similar pattern. Such behavior is nicely reflected in the values of skewness, which turns out to be around $0.5$ for all the QC measures. Since previous studies show that random $N$-party pure states, with $N > 11$ are highly entangled \cite{winter, eisert}, it will be interesting to see  the distribution of the monogamy scores for higher number of parties. In this respect, let us first consider the monogamy score in an asymptotic scenario.

%\sout{In the previous section, we notice that almost all measures become monogamous when squared. We now probe further into the squared scores of those that became monogamous when squared while varying the number of parties. When the number of parties is very large, the monogamy score for almost all measures tends to unity.}

\textbf{Theorem 1.} \emph{For a random pure state of N qubits, the monogamy score for any entanglement measure $\mathcal{Q}$ approaches to unity with the increase of the number of parties provided the entanglement measure reduces to von Neumann entropy of the local density matrix for pure states.}
\begin{proof}
For pure states, of $N$ parties, $|\psi\rangle_{123...N}$, the first term in the monogamy score defined in Eq. \eqref{eq:def_ms}, $\mathcal{Q}_{1:\text{rest}}$, for a large variety of entanglement measures, reduces to the von Neumann entropy of the reduced density matrix, $S(\rho_1) = -\text{tr}(\rho_1 \log_2 \rho_1)$, where $\rho_1 = \text{tr}_{23...N}(|\psi\rangle_{123...N}\langle\psi|)$. 
%Now following the proofs given in \cite{entropy1,entropy2,entropy3}, 
For a state with $N$ qubits, the mean entropy, averaged over the Haar measure, reads \cite{entropy1,entropy2,entropy3} 
\begin{eqnarray}
\langle S(\rho_1)\rangle = \log_2 e \Big(\sum_{2^{N-1} +1}^{2^{N}}\frac{1}{j} - \frac{1}{2^{N}} \Big) \approx 1 - \frac{\log_2 e}{2^{N-1}},
\end{eqnarray}
%The second term in the above expression vanishes for large $n$, and we have 
so that for $N \rightarrow \infty$, we have
\begin{eqnarray}
\langle \mathcal{Q}_{1:\text{rest}} \rangle = \langle S(\rho_1)\rangle_{N\rightarrow\infty} \rightarrow 1.
\end{eqnarray}
Although we have taken the first party to be the nodal observer, the results remain unchanged for other nodal observers as well.
Therefore almost all states, $\mathcal{Q}_{1:\text{rest}} = 1$ for large $N$.
% here, we could have taken any party to be the nodal observer and all results would have been exactly same.

Let us now try to put bounds on the bipartite QC, $Q_{1:i}$ in Eq. \eqref{eq:def_ms}. If we restrict ourselves to spin-$1/2$ particles, all $\rho_{1:i}$ are two-qubit states having rank 2. Moreover, we know \cite{ppt1,ppt2} that the partial transposition criterion is necessary and sufficient for entanglement in this case. All bipartite states obtained from an $N$ qubit random pure state will have positive partial transpose \cite{entropy_bounds} for $N > 6 + \log_2 \frac{11}{16}$, and hence are separable, having vanishing entanglement, i.e.
%\textcolor{red}{$\mathcal{Q}_{1:i} \rightarrow 0$.}
%Since we are working with qubit systems, the PPT (Positive Partial Transpose) criterion \cite{ppt} is both necessary and sufficient. From \cite{entropy_bounds} we conclude that all bipartite reductions $\rho_{1:i}$ of $|\psi\rangle_{123...n}$ are PPT on average when
%\begin{eqnarray}
%n > 6 + \log_2 \frac{11}{16}.
%\end{eqnarray} 
 we get $ \mathcal{Q}_{1:i} $ identically vanishing for sufficiently large $N$, for almost all states. Therefore, for the  entanglement measures considered, for sufficiently large values of $N$, 
\begin{eqnarray}
\delta_\mathcal{Q} = 1,
\end{eqnarray}
for almost all $N$-qubit states. This completes the proof.
\end{proof}

\textbf{Remark:} Numerical simulations of four, five and six-qubit states unveils that $\langle \mathcal{Q}_{1:i} \rangle$ approaches its extremal value, i.e., $\langle \mathcal{Q}_{1:i} \rangle \rightarrow 0$, faster than the way $\langle \mathcal{Q}_{1:\text{rest}} \rangle$ approaches unity with the increase of $N$.

Numerical simulations reveal that not only for entanglement measures, but also for information-theoretic QC measures, for high $N$, say for random six-qubit states, $\mathcal{Q}(\rho_{1:i}) \rightarrow 0$ and $\langle \delta_\mathcal{Q} \rangle \rightarrow 1$ as $\langle \mathcal{Q}_{1:\text{rest}} \rangle \rightarrow 1$, which is in agreement with the asymptotic case.
% irrespective of the choice of $\mathcal{Q}$. The above results show the limit of the monogamy score in the asymptotic case. 
Let us now describe the way by which it approaches to unity for random multiparty pure states (cf. \cite{eisert}).
%\textcolor{red}{independent of entanglement} of $\rho_{1:i}$ of random pure states can be non-vanishing. However, as we show now, 

%Now we investigate how the transition to unit monogamy score takes place numerically, by increasing the number of parties.
% from $3$ to $6$.
We again Haar uniformly \cite{Karolbook} generate random pure states of three-, four-, five- and six-qubits and compute $\delta_{\mathcal{Q}^2}$. To analyze the data in a quantitative manner, we evaluate mean, variance and skewness for various measures as presented in Table. \ref{table:stat_36}. Figs. \ref{fig:ConcSqParties} and \ref{fig:DiscSqParties} highlight the change in the distribution of $\delta_{\mathcal{C}^2}$ and $\delta_{\mathcal{D}_\leftarrow^2}$ with the variation of number of qubits. Analysis of the Tables \ref{table:stat_36} and \ref{table:mean_3-6}, and Figs.  \ref{fig:ConcSqParties} and \ref{fig:DiscSqParties} reveal certain aspects of monogamy scores --

\begin{enumerate}
%\item As expected, when the number of parties are increased, there is a shift of the average monogamy score towards unity. When the number of parties are incresed to $4$ from $3$, apart from a higher mean monogamy score, for all measures, the distribution makes a transition from being right skewed to left skewed. The left skewness in the distribution of monogamy score persists for $n=5,6$ also.
\item Means of $\delta_{\mathcal{Q}^2}$ increases with $N$ $\forall \mathcal{Q}$ and becomes more than $0.92$ for $N=6$. Moreover, entanglement measures approach unity faster than the information theoretic one.

\item All the QC measures make a transition from being right skewed  to being skewed to the left  with $N \geq 4$.

%\item For $n=6$, the monogamy score of a random pure state is greater than $0.9$ irrespective of the measure of quantum correlation, i.e., $\langle \delta_\mathcal{Q}\rangle|_{n=6}>0.9$.
%
%\item The distribution of monogamy score becomes progressively sharper and concentrated near the mean value with the increase in $n$.

%\item We also notice that on increasing $n$, $\delta_\mathcal{Q}(|\psi_{123...n}\rangle)$ tends to the maximal value and the sum over all two-party reduced density matrices tends to $0$, independent of the measure chosen. Therefore, this shift is a general property of any correlation measure and not of a specific measure or type of measure. 
\end{enumerate}

%The monogamy score for any quantum correlation measure $\mathcal{Q}$, approaches unity with the increase in the number of qubits.
%\end{theorem}
%  
Theorem 1 coupled with the numerical simulations gives a comprehensive picture of how the distributions of monogamy score approaches unit mean with the increase of $N$. Therefore, almost all the random pure states not only satisfy monogamy relation for any arbitrary squared QC measure, $\mathcal{Q}^2$, but it possess a very high value of the same with states having moderate number of parties. For example, among $10^6$ random six-qubit pure states, $91.34 \%$ states have $\delta_{\mathcal{C}^2}$ more than $0.9$.

The geometric measure of entanglement \cite{geoent}, defined as $E_g(|\psi\rangle) = -\log_2\sup_{\alpha \in \mathcal{P}}|\langle\alpha|\psi\rangle|^2$ with $\mathcal{P}$ being the set of all product states, rises rapidly with the increase in the number of parties.  Specifically, it was shown that the fraction of states having $E_g \leq N-2\log_2(N) - 3$ is smaller than $e^{-N^2}$. 
Our results which shows how the maximal value is attained by increasing the number of parties have a striking resemblance with the above result. Moreover, this similarity is independent of the measure used to construct the monogamy scores. Thus, we can safely assert that on increasing the number of parties, almost all randomly chosen pure states have maximal monogamy score and that the state possess high QC.

\begin{widetext}
\begin{center}
\begin{table*}[h]
\centering
\begin{tabular}{|c|c|c|c|c|c|c|}
\hline
{$X = \mu, \sigma^2, \kappa$}  & \multicolumn{2}{|c|}{\textbf{Mean}} & \multicolumn{2}{|c|}{\textbf{Variance}} & \multicolumn{2}{|c|}{\textbf{Skewness}}  \\\hline
\textbf{\backslashbox{$\mathcal{Q}^2 \downarrow$}{$N$ $\rightarrow$}} & 3 & 6 & 3 & 6 & 3 & 6 \\\hline
$\mathcal{C}^2$ & 0.33335 & 0.95732 & 0.034293 & 0.0013211 & 0.50371 & -1.4482 \\\hline
$\mathcal{N}^2$ & 0.41329 & 0.95373 & 0.027109 & 0.0013209 & 0.19832 & -1.4483 \\\hline
$EoF^2$ & 0.3985 & 0.93432 & 0.040395 & 0.00259 & 0.32512 & -1.3846 \\\hline
$\mathcal{L}^2$ & 0.37581 & 0.96605 & 0.023849 & 0.00073 & 0.040478 & -1.5107\\\hline
$\mathcal{D}^2_\leftarrow$ & 0.42308 & 0.92927 & 0.040715 & 0.0025693 & 0.23827 & -1.3762\\\hline
\end{tabular}
\caption{Statistical quantities ($X = \mu, \sigma^2, \kappa$) of $\delta_{\mathcal{Q}^2}$ for three- and six-qubit states.}
\label{table:stat_36}
\end{table*}
\end{center}
\end{widetext}

\begin{table}[h]
\centering
\begin{tabular}{|c|c|c|c|c|}
\hline
\textbf{$X = \mu$} & \multicolumn{4}{|c|}{\textbf{Mean}} \\\hline
\textbf{\backslashbox{$\mathcal{Q}^2 \downarrow$}{$N$ $\rightarrow$}} & 3 & 4 & 5 & 6\\\hline
$\mathcal{C}^2$ & 0.33335 & 0.72941 & 0.90175 & 0.95732\\\hline
$\mathcal{N}^2$ & 0.41329 & 0.75322 & 0.90303 & 0.95373\\\hline
$EoF^2$ & 0.3985 & 0.73239 & 0.87166 & 0.93432\\\hline
$\mathcal{L}^2$ & 0.37581 & 0.75106 & 0.92158 & 0.96605\\\hline
$\mathcal{D}^2_\leftarrow$ & 0.42308 & 0.70413 & 0.85582 & 0.92927\\\hline
\end{tabular}
\caption{Mean of $\delta_{\mathcal{Q}^2}$ for $N = 3$ to $6$ parties.}
\label{table:mean_3-6}
\end{table}

Furthermore, it was shown \cite{Dur_SLOCC} that monogamy scores with squared concurrence, $\delta_{\mathcal{C}^2}$, also known as three tangle is an entanglement monotone, and consequently all its positive powers are also valid entanglement measures. Although this is no longer true for monogamy scores based on other QC measures.
%, although we notice that the distribution of monogamy scores constructed using all QC measures possess identical features. We can safely conclude that if the monogamy power is greater than the critical value ($\alpha > \alpha_c$), thereby guaranteeing the non-negativity of the monogamy score, then, independent of $\mathcal{Q}$, the monogamy scores can be argued as good as multiparty entanglement.
Analyzing our data, we can conclude that if the monogamy scores are properly written, so that non-negativity is guaranteed, then their distributions mimic the pattern of multiparty entanglement for moderate number of parties.
Moreover, our results also establish a connection between the restrictions in sharability of bipartitie QC in a multiparty quantum state and the amount of multiparty QC contained in the state.

%Coming back to the measures themselves, among the measures considered in the manuscript, only concurrence and entanglement of formation and consequently all of their positive powers are entanglement monotones. However, 

\section{Conclusion}
\label{sec:conclusion}
For multiparty quantum states, the knowledge about the distribution of quantum correlation among its various parties lies at the heart of the characterization of multiparty entanglement or more generally, multiparty quantum correlations (QC). Unlike classical correlations, quantum mechanics puts restriction on arbitrary sharing of QC. Such constraints are encapsulated in the concept of monogamy and is quantified in terms of the \emph{monogamy score}.

The monogamy score can be defined for any valid quantum correlation measure and the same raised to any non-negative power. In this paper, we compute the critical monogamy power above which all the random pure states becomes monogamous for several paradigmatic QC measures. We also show that it decreases with the increase in number of parties. We find that the fraction of nonmonogamous states remains frozen for certain powers of QC measure in a monogamy score.

Both analytically and numerically, we report that the mean of the distribution  of monogamy scores for large number of parties of random pure multiqubit states approaches to unity irrespective of the QC measure. Our analysis unveils an universal character, independent of the measure, in the distribution of monogamy scores and its statistically relevent quantities like mean, variance and skewness.

From the distribution of monogamy score,  it is clear that almost all random pure states having moderate number of parties are not only monogamous with respect to any QC measure, but possess high value of the corresponding monogamy score. 
Such results can have some important applications towards building quantum cryptography involving multiple parties. Moreover, 
the observations in behavior of monogamy scores are in good agreement with the known results obtained for geometric measure of entanglement for random multiparty states. Additionally, our analysis of mean and skewness of the distributions show the path in which the distributions change with the variation of the number of parties. 
%We believe that our results can have important implications in quantum cryptography networks.

\begin{acknowledgments}
%We thank Titas Chanda for discussions. 
%AK acknowledges research fellowship from the Department of Atomic Energy, Government of India. 
This research was supported in part by the 'INFOSYS scholarship for senior students'. Numerical results have been obtained using the Quantum Information and Computation library (QIClib) (\url{https://titaschanda.github.io/QIClib}). TC was supported in part by Quantera QTFLAG project 2017/25/Z/ST2/03029 of National Science Centre (Poland). Soorya Rethinasamy acknowledges the hospitality of Harish-Chandra Research Institute.
\end{acknowledgments}

\appendix
\section{}
\label{appendix}
We describe the parameterizations used to describe the two SLOCC inequivalent classes of three-qubit pure states. Furthermore, we  provide definitions of the quantum correlation measures, used in the main text.

\subsection{GHZ- and W-class states}
Any arbitrary three-qubit random pure state can be represented as
$$|\psi\rangle = \sum\limits_{i=1}^8 \alpha_i |i_1 i_2 i_3\rangle,$$
where $\alpha_i = \alpha'_i + i \alpha''_i$, with $\alpha'
_i$ and $\alpha''_i$ being chosen randomly from a normal distribution with mean $0$ and variance unity and $|i_k\rangle = |0\rangle$ or $|1\rangle$, $k=1, 2, 3$. Any random pure three-qubit state belongs to the GHZ-class, having $\delta_{\mathcal{C}^2}>0$

On the other hand, another important class of three-qubit pure states which are SLOCC-inequivalent with the GHZ-class is the W-class states, given by
$$|\psi_W\rangle = \sqrt{a}|001\rangle + \sqrt{b}|010\rangle + \sqrt{c}|100\rangle + \sqrt{d}|000\rangle$$ upto local unitary operations. Since among three-qubit pure states, the W-class states belong to the set of measure zero, simulation of random three-qubit pure states do not produce states from the W-class. Since it is an important class and it's trends of the distribution can be important, we numerically simulate the W-class states by randomly choosing parameters from normal distribution.

\subsection{Entanglement Measures}
Some typical measures from the entanglement separability paradigm are discussed below.
For a pure bipartite quantum state, $|\psi_{AB}\rangle$, entanglement is uniquely defined as the von Neumann entropy of its reduced density matrices, given by
\begin{eqnarray}
\mathcal{E}(|\psi_{AB}\rangle) = \mathcal{S}(\rho_A) = \mathcal{S}(\rho_B)
\end{eqnarray}
where $\mathcal{S}(\sigma) \equiv -tr(\sigma log \sigma)$, and $\rho_A$ and $\rho_B$ are the local density matrices of $|\psi_{AB}\rangle$.
For mixed bipartite states, there are more than one inequivalent quantifiers of entanglemnt. The ones considered in this manuscript are the known computable ones.
\begin{enumerate}
 \item The \emph{entanglement of formation} \cite{eof} for a mixed state $\rho_{AB}$ is defined as
\begin{eqnarray}
EoF(\rho_{AB}) = \min\limits_{\{p_i, |\psi^i_{AB}\rangle\}} \sum\limits_i p_i \mathcal{E}(|\psi_{AB}^i\rangle),
\end{eqnarray}
where the minimization is taken over all possible pure state decompositions, $\rho_{AB} = \sum_i p_i |\psi^i_{AB}\rangle\langle\psi^i_{AB}|$. It can be computed for arbitrary two-qubit states, explained below.

\item The \emph{concurrence} \cite{concurrence} for a pure state, $|\psi_{AB}\rangle$, is defined as 
\begin{eqnarray}
\mathcal{C}(\psi_{AB}) = |\langle\psi_{AB}|\tilde{\psi}_{AB}\rangle|,
\end{eqnarray}
where $|\tilde{\psi}_{AB}\rangle = \sigma_y \otimes \sigma_y |\psi^*_{AB}\rangle$ with $|\psi^*_{AB}\rangle$ being the complex conjugate of $|\psi_{AB}\rangle$ in the computational basis $\{|00\rangle, |01\rangle, |10\rangle, |11\rangle\}$.

For a general two-qubit density matrix, $\rho_{AB}$, we first define the spin flipped density matrix, $\tilde{\rho}_{AB} = (\sigma_y \otimes \sigma_y) \rho^*_{AB}(\sigma_y \otimes \sigma_y)$, and the operator $\mathcal{R} = \sqrt{\sqrt{\rho_{AB}}\rho^*_{AB}\sqrt{\rho_{AB}}}$. The definition of concurrence then reduces to $\mathcal{C}(\rho_{AB}) = \min\{0, \lambda_1 - \lambda_2 - \lambda_3 - \lambda_4\}$, where $\lambda_i$s are the eigenvalues of $R$ in decreasing order. The entanglement of formation is then given by $EoF(\rho_{AB}) = \mathcal{F}(\mathcal{C}(\rho_{AB}))$ where $\mathcal{F}(\mathcal{C})$ reads as
\begin{eqnarray}
\mathcal{F}(\mathcal{C}) = h\left(\frac{1+\sqrt{1-\mathcal{C}^2}}{2}\right),
\end{eqnarray}
with $h(x) = -x\log_2(x) - (1-x)\log_2(1-x)$ being the well known binary entropy.

\item \emph{Negativity} \cite{neg} is defined as the twice of the absolute sum of negative eigenvalues of the partially transposed density matrix, $\rho^{T_B}_{AB}$, where partial transposition is taken with respect to B.
Mathematically,
\begin{eqnarray}
\mathcal{N}(\rho_{AB}) = ||\rho^{T_B}_{AB}|| - 1 = ||\rho^{T_A}_{AB}|| - 1, 
\end{eqnarray}
where $||A|| = tr\sqrt{A^\dagger A}$. 

Note that we multiply by 2 to make its maximum for two-qubit maximally entangled states to be unity.
\item The \emph{logarithmic negativity} \cite{neg,logneg} for a bipartite state $\rho_{AB}$ is defined in terms of negativity as 
\begin{eqnarray}
\mathcal{L}(\rho_{AB}) = \log_2(\mathcal{N}(\rho_{AB}) + 1).
\end{eqnarray}
\end{enumerate}
positivity of $\mathcal{L}(\rho_{AB})$ guarantees that the state is entangled and distillable.

\subsection{QC Measures Independent of Entanglement}
Some representative quantum correlation measures from the information-theoretic paradigm are given below.

\begin{enumerate}
\item \emph{Quantum discord}: 

Uninterrogated quantum conditional entropy of a bipartite state $\rho_{AB}$, shared between $A$ and $B$, is defined as $\tilde{\mathcal{S}}(\rho_{A|B}) = \mathcal{S}(\rho_{AB}) - \mathcal{S}(\rho_B)$. Similarly, the interrogated conditional entropy is defined as 
\begin{eqnarray}
\mathcal{S}(\rho_{A|B}) = \min\limits_{\Pi^B_i} \sum\limits_i p_i \mathcal{S}(\rho_{A|i}),
\end{eqnarray}
 where the minimization is performed over all complete sets of projective measurements with
\begin{eqnarray}
\rho_{A|i} = Tr_B[(\mathbb{I}_A \otimes \Pi^B_i)\rho_{AB}(\mathbb{I}_A \otimes \Pi^B_i)]/p_i,
\end{eqnarray}
where $p_i = Tr_{AB}[(\mathbb{I}_A \otimes \Pi^B_i)\rho_{AB}(\mathbb{I}_A \otimes \Pi^B_i)]$.

Hence, uninterrogated quantum mutual information can be written as
\begin{eqnarray}
\tilde{I}(\rho_{AB}) = \mathcal{S}(\rho_A) - \tilde{\mathcal{S}}(\rho_{A|B}),
\end{eqnarray}
while the interrogated quantum mutual information reads as
\begin{eqnarray}
I_\leftarrow(\rho_{AB}) = \mathcal{S}(\rho_A) - \mathcal{S}(\rho_{A|B}),
\end{eqnarray}
where the symbol "$\leftarrow$" in the subscript indicates that the measurement is performed on the second subsystem, B.

Quantum discord \cite{discord1,discord2} is defined as the difference between uninterrogated and interrogated quantum mutual information, given by
\begin{eqnarray}
\mathcal{D}_\leftarrow(\rho_{AB}) = \tilde{I}(\rho_{AB}) - I_\leftarrow(\rho_{AB}),
\end{eqnarray} 
which can be called quantum discord (right).
Naturally, when measurement is performed in the first party, we get quantum discord (left), denoted by $\mathcal{D}_\rightarrow$.
Throughout the paper, unless mentioned, we compute $\mathcal{D}_\leftarrow$. The results remain qualitatively unchanged for $\mathcal{D}_\rightarrow$.

\item \emph{Quantum work deficit}:

The amount of extractable pure states under a set of global operations,  called``closed operations" (CO) is given by
\begin{eqnarray}
I_{CO} = \log_2 \dim(\mathcal{H}) - \mathcal{S}(\rho_{AB}).
\end{eqnarray}
On the other hand, the amount of extractable pure states under a set of closed local operations and classical communication (CLOCC) can be 
\begin{eqnarray}
I_{CLOCC} = \log_2 \dim(\mathcal{H}) - \min \mathcal{S}(\rho'_{AB})
\end{eqnarray}	
with minimizaion being performed over all local projective measurements on the second party and where $\rho'_{AB} = \sum p_i \rho_{A|i}$. Quantum work deficit \cite{wd1,wd2} can be argued as a measure of QC and is defined to the residual work extraction by CO after extraction by CLOCC and can be represented as
\begin{eqnarray}
\mathcal{W}_\leftarrow(\rho_{AB}) = I_{CO} - I_{CLOCC}.
\end{eqnarray}
Like quantum discord, when the measurement is performed in the first party, we have $\mathcal{W}_\rightarrow$.
\end{enumerate}

 %\label{appendix}

%An arbitrary three qubit state from the GHZ class can be parameterized \cite{Dur_SLOCC} as
%\begin{eqnarray}
%|\psi_{GHZ}\rangle &=& \sqrt{N}\big (c_\delta |0\rangle|0\rangle|0\rangle + s_\delta e^{i\theta}|\phi_A\rangle|\phi_C\rangle|\phi_B\rangle \big),  \nonumber 
%\end{eqnarray}
%where 
%\begin{eqnarray}
%|\phi_A\rangle &=& c_\alpha |0\rangle +s_\alpha |1\rangle \nonumber \\
%|\phi_B\rangle &=& c_\beta |0\rangle +s_\beta |1\rangle \nonumber \\
%|\phi_C\rangle &=& c_\gamma |0\rangle +s_\gamma |1\rangle \nonumber 
%\end{eqnarray}
%and $N = (1 + 2c_\delta s_\delta c_\alpha c_\beta c_\gamma c_\theta)^{-1} \in (\frac{1}{2},\infty) $ is the normalisation factor. The ranges for the five parameters are $\delta \in (0,\frac{\pi}{4}],\alpha,\beta,\gamma \in (0,\frac{\pi}{2}] \text{ and } \theta \in (0,2\pi]$. Here $c_x \text{ and } s_x$ represents $\cos x \text{ and } \sin x$ respectively. Similarly, an arbitrary three-qubit state belonging to the W class, upto local unitary operations, can be represented as 
%\begin{eqnarray}
%|\psi_W\rangle = \sqrt{a}|001\rangle + \sqrt{b}|010\rangle + \sqrt{c}|001\rangle + \sqrt{d}|000\rangle, \nonumber 
%\end{eqnarray}
%%We randomly generate pure tripartite states Haar uniformly as described in Ref. \cite{Karolbook} to investigate Figs.  \ref{fig:Histo_neg_m} and \ref{fig:Histo_neg_m_2copy}. 

%\begin{thebibliography}{19}


\begin{thebibliography}{19}
\bibitem{othernodal} Other parties  of a multipartite state can also serve as nodal observer.
\end{thebibliography}
\end{document}